\documentclass[fleqn,usenatbib]{mnras}

\usepackage{newtxtext,newtxmath}
\usepackage[T1]{fontenc}
\usepackage{array}
\usepackage{booktabs}
\usepackage{multirow}
\usepackage{hyperref}
\usepackage{orcidlink}
\DeclareRobustCommand{\VAN}[3]{#2}
\let\VANthebibliography\thebibliography
\def\thebibliography{\DeclareRobustCommand{\VAN}[3]{##3}\VANthebibliography}

\usepackage{graphicx}	% Including figure files
\usepackage{amsmath}	% Advanced maths commands
\usepackage{xspace}
\newcommand{\igr}{IGR~J17091$-$3624\xspace} 
\newcommand{\grs}{GRS~1915$+$105\xspace}

\newcommand{\ixpe}{\textit{IXPE}\xspace} 
\newcommand{\integral}{\textit{INTEGRAL}\xspace} 
\newcommand{\nustar}{\textit{NuSTAR}\xspace} 
\title[X-ray polarisation of \igr]{The very high X-ray polarisation of accreting black hole IGR J17091$-$3624 in the hard state}

\author[M. D. Ewing et al.]{Melissa Ewing\orcidlink{0000-0001-9349-8271},$^{1}$\thanks{E-mail: publications@ras.ac.uk (KTS)}
Maxime Parra\orcidlink{0009-0003-8610-853X},$^{2}$
Guglielmo Mastroserio\orcidlink{0000-0003-4216-7936},$^{3}$
Alexandra Veledina\orcidlink{0000-0002-5767-7253},$^{4,5}$
Adam Ingram\orcidlink{0000-0002-5311-9078},$^{1}$
\newauthor 
Michal Dov\v{c}iak\orcidlink{0000-0003-0079-1239},$^{6}$
Javier A. Garc\'ia\orcidlink{0000-0003-3828-2448},$^{7,8}$
Thomas D. Russell\orcidlink{0000-0002-7930-2276},$^{9}$
Maria C. Baglio\orcidlink{0000-0003-1285-4057},$^{10}$ 
Juri Poutanen\orcidlink{0000-0002-0983-0049},$^{4}$
\newauthor 
Oluwashina Adegoke,$^{8}$
Stefano~Bianchi\orcidlink{0000-0002-4622-4240},$^{11}$
Fiamma Capitanio\orcidlink{0000-0002-6384-3027},$^{12}$
Riley Connors,$^{13}$
Melania Del Santo\orcidlink{0000-0002-1793-1050},$^{9}$
\newauthor 
Barbara De Marco\orcidlink{0000-0003-2743-6632},$^{14}$
 Mar{\'i}a D{\'i}az Trigo\orcidlink{0000-0001-7796-4279}, $^{15}$
Poshak Gandhi,$^{16}$
Maitrayee Gupta\orcidlink{0000-0003-0976-8932},$^{6}$
Chulsoo Kang,$^{2}$
\newauthor
Elias Kammoun\orcidlink{0000-0002-0273-218X},$^{8}$ 
Vladislav~Loktev\orcidlink{0000-0001-6894-871X},$^{4,17}$
Lorenzo~Marra\orcidlink{0009-0001-4644-194X},$^{12}$
Giorgio Matt\orcidlink{0000-0002-2152-0916},$^{11}$
Edward~Nathan,$^{8}$ \orcidlink{0000-0002-9633-9193}
\newauthor 
Pierre-Olivier Petrucci\orcidlink{0000-0001-6061-3480},$^{18}$
Megumi Shidatsu\orcidlink{0000-0001-8195-6546},$^{2}$
James F. Steiner\orcidlink{0000-0002-5872-6061},$^{19}$
Francesco Tombesi\orcidlink{0000-0002-6562-8654},$^{20, 21, 22}$
\newauthor 
and Federico M. Vincentelli,$^{16}$
\\
$^{1}$School of Mathematics, Statistics, and Physics, Newcastle University, Newcastle upon Tyne, NE1 7RU, UK\\
$^{2}$Department of Physics, Ehime University, 2-5, Bunkyocho, Matsuyama, Ehime 790-8577, Japan\\
$^{3}$Dipartimento di Fisica, Universit\`{a} degli Studi di Milano, Via Celoria 16, I-20133 Milano, Italy\\
$^{4}$Department of Physics and Astronomy, 20014 University of Turku, Finland\\
$^{5}$Nordita, KTH Royal Institute of Technology and Stockholm University, Hannes Alfv\'ens v\"ag 12, SE-10691 Stockholm, Sweden\\
$^{6}$Astronomical Institute of the Czech Academy of Sciences, Bo\v{c}n\'{i} II 1401, 14100 Praha, Czech Republic; \\
$^{7}$ NASA Goddard Space Flight Center, Greenbelt, MD 20771, USA\\
$^{8}$Cahill Center for Astrophysics, California Institute of Technology, 1216 East California Boulevard, Pasadena, CA 91125, USA \\
$^{9}$INAF, Istituto di Astrofisica Spaziale e Fisica Cosmica, Via U. La Malfa 153, I-90146 Palermo, Italy\\
$^{10}$INAF--Osservatorio Astronomico di Brera, Via Bianchi 46, I-23807 Merate (LC), Italy\\
$^{11}$ Dipartimento di Matematica e Fisica, Università degli Studi Roma Tre, Via della Vasca Navale 84, 00146 Roma, Italy\\
$^{12}$ INAF - IAPS, via del fosso del Cavaliere 100, 00133 Roma, Italy\\
$^{13}$ Department of Physics, Villanova University, 800 E. Lancaster Avenue, Villanova, PA 19085, USA\\
$^{14}$ Departament de Fis\'{i}ca, EEBE, Universitat Polit\`ecnica de Catalunya, Av. Eduard Maristany 16, 08019 Barcelona, Spain\\
$^{15}$ ESO, Karl-Schwarzschild-Strasse 2, 85748, Garching bei M{\"u}nchen, Germany\\
$^{16}$ School of Physics \& Astronomy, University of Southampton, Southampton SO17 1BJ, UK\\
$^{17}$Department of Physics, P.O. Box 64, 00014 University of Helsinki, Finland; \\
$^{18}$Universit\'{e} Grenoble Alpes, CNRS, IPAG, 38000 Grenoble, France\\
$^{19}$ Center for Astrophysics \textbar\ Harvard \& Smithsonian, 60 Garden St, Cambridge, MA 02138, USA\\
$^{20}$ Physics Department, Tor Vergata University of Rome, Via della Ricerca Scientifica 1, 00133 Rome, Italy\\
$^{21}$ INAF – Astronomical Observatory of Rome, Via Frascati 33, 00040 Monte Porzio Catone, Italy\\
$^{22}$ INFN - Rome Tor Vergata, Via della Ricerca Scientifica 1, 00133 Rome, Italy\\
}

\date{Accepted XXX. Received YYY; in original form ZZZ}

\pubyear{\the\year{}}

\begin{document}
\label{firstpage}
\pagerange{\pageref{firstpage}--\pageref{lastpage}}
\maketitle

\begin{abstract}
We report the first detection of the X-ray polarisation of the transient black hole X-ray binary \igr taken with the \textit{Imaging X-ray polarimetry Explorer} (\ixpe) in March 2025, and present the results of an X-ray spectro-polarimetric analysis. The polarisation was measured in the 2--8 keV band with 5.2$\sigma$ statistical confidence. We report a polarisation degree (PD) of $9.1\pm1.6$ per cent and a polarisation angle of $83\degr\pm5\degr$ (errors are $1\sigma$ confidence). There is a hint of a positive correlation of PD with energy that is not statistically significant. We report that the source is in the corona-dominated hard state, which is confirmed by a hard power-law dominated spectrum with weak reflection features and the presence of a Type-C quasi-periodic oscillation at $\sim0.2$~Hz. The orientation of the emitted radio jet is not known, and so we are unable to compare it with the direction of X-ray polarization, but we predict the two to be parallel if the geometry is similar to that in Cygnus X-1 and Swift J1727.8-1613, the two hard state black hole binaries previously observed by \ixpe. In the Comptonisation scenario, the high observed PD requires a very favourable geometry of the corona, a high inclination angle (supported by the presence of a dip in the light curve) and possibly a mildly relativistic outflow and/or scattering in an optically thick wind.
\end{abstract}


%%%%%%%%%%%%%%%%%%%%%%%%%%%%%%%%%%%%%%%%%%%%%%%%%%

%%%%%%%%%%%%%%%%% BODY OF PAPER %%%%%%%%%%%%%%%%%%

\section{Introduction}
A black hole (BH) X-ray binary (XRB) is a system whereby a stellar-mass BH accretes matter from a companion star, producing an extremely high X-ray flux. They are categorised into {transient} or {persistent} systems, where transient systems exhibit outbursts followed by quiescence and persistent systems maintain a steady flux above a quiescent level. In both cases we observe spectral state transitions \citep{Fender2004,Done2007,Belloni2010}. In transient systems, the outburst begins in the hard state, which is dominated by the Comptonisation of photons from a quasi-thermal accretion disc \citep{shakura1973black,Novikov1973} or internal  synchrotron photons \citep{Poutanen2009,MalzacBelmont2009,Veledina2011} in a hot cloud of electrons known as the `corona' \citep{Sunyaev1985}. A steady compact jet is launched from the system during the hard state. The system then transitions through the intermediate state, where both the disc and corona are prominently observed in the spectrum,
% the source moves
to the
% high
soft state, where now the spectrum is dominated by the disc and the compact jet is quenched. Again transitioning through intermediate states, the source moves back to the hard state and then quiescence. In the case of persistent sources, we observe these spectral transitions without the flux dropping to quiescence. 

Because XRBs cannot be spatially resolved, the geometry and position with respect to the disc of the hard X-ray corona is still a matter of debate \citep{Poutanen2018}. Several models have been proposed such as the `sandwich' model, whereby the corona is powered by energy dissipation above and below the disc \citep{Haardt1993,Svensson1994}, 
the `magnetic flare' model \citep{Galeev1979,Beloborodov1999} with the localised energy dissipation regions above the disc, the `jet-base' model \citep{Markoff2005} in which the corona is vertically extended at the base of the jet, the `lamppost' model \citep{matt1993b}
%\citep{martocchia2002origin} 
where a compact corona lies above the BH on its spin axis, and the `truncated disc' model \citep{eardley1975cygnus,Esin1997,Poutanen1997} whereby the disc evaporates beyond a truncation radius into a large scale-height hot flow. Each of these models can produce the same spectra, and so it is vital that we collect more information on these systems in order to break this degeneracy. 

With the launch of the \textit{Imaging X-ray Polarimetry Explorer} \citep[\ixpe;][]{Weisskopf2022}, we now have access to linear X-ray polarisation information in the 2--8 keV energy range. We can measure the polarisation degree (PD) -- the extent to which the source X-ray photons have aligned electromagnetic fields -- and the polarisation angle (PA) -- the angle at which the electromagnetic fields align. The PD increases with asymmetry of the emission region; i.e. with larger (more edge-on) inclination and smaller aspect ratio \citep{Schnittman2010,Ursini2022,Poutanen_2023}. For Comptonised radiation, the PA aligns perpendicular to the major axis of the scattering region (i.e. the corona). This happens because photons scatter preferentially in the plane perpendicular to their polarisation vector, and they experience more scatterings along the major axis due to its higher scattering optical depth \citep{Sunyaev1985,Ursini2022,ingram2023xraypolarisationseyfert1}. 

%Since hard coronal emission is driven by Comptonisation, polarisation reveals the corona's asymmetry and position in the sky. In a symmetric corona (e.g., a sphere), polarisation vectors cancel out, leading to 0\% net PD. In an asymmetric corona, this symmetry breaks, with asymmetry increasing as the region becomes more inclined to our view \citep{Schnittman2010}. PD increases with greater asymmetry since photons along the major axis are more abundant due to higher scattering optical depth, and scattering polarises photons perpendicular to their electric field vector, aligning the PA perpendicular to the major axis plane.\\
%If we know the inclination of a source, then we can make theoretical predictions on the expected PD.

The first BH XRB observed by \ixpe during a hard state was Cygnus~X-1 (hereafter Cyg X-1), the brightest  persistent BH XRB observed in the sky. Linear polarisation was detected with a $>20~\sigma$ statistical confidence, with PD = $4.0\pm 0.2$ per cent and PA = $-20\fdg7\pm 1\fdg4$ \citep{Krawczynski_2022}. The PA was found to align with the direction of the radio jet,  suggesting that the corona is perpendicular to the jet and elongated in the disc plane. This favours horizontally extended models such as the truncated disc model over those vertically extended, such as the lamppost or jet-base model. With a relatively low inclination angle of $i=27\fdg5\pm 0\fdg8$ inferred from optical observations of the binary \citep{Miller2021}, a low PD of $\approx 1$~per cent was expected \citep{Krawczynski_2022_lowpol}. 
One explanation of this high observed PD is that the corona is more inclined than the binary system. This could be due to a misalignment between the BH and binary spin axes giving rise to a warped disc.
% have occurred during the BH formation where the BH spin axis was initially misaligned with the binary plane \citep{Bardeen1975,Liska2019}.
%could possibly occur if the BH was produced with a misalinged spin axis.
Another potential explanation is an outflowing corona; \cite{Poutanen_2023} showed that a mildly relativistic bulk coronal electron velocity of $v \sim 0.4~c$ can boost the PD to the observed value for an aligned system \citep[see also][]{Dexter2024}.
If the former model is true, then this suggests that a significant subset of observed systems will exhibit a low PD. If the latter is true, then this implies that a high PD is common in XRB systems observed in the hard state. 

The second BH XRB  observed by \ixpe in the hard state was Swift J1727.8$-$1613, a transient source that was discovered after going into a bright outburst in 2023 \citep{Veledina2023}. Again it was found that the PD was $\approx 4$~per cent and that the PA was aligned with the orientation of the radio jet \citep{Wood2024}. This outburst was the first time that X-ray polarisation had been tracked across a hard to soft state transition, where the PD was found to slowly decrease across the transition and the PA stayed constant \citep{ingram2024trackingxraypolarizationblack}. The PD was later seen to reduce dramatically in the soft state \citep{Svoboda2024}, before recovering to the same level as before in the reverse soft to hard state transition \citep{Podgorny2024}.

%To determine whether the similarities between these sources is significant or coincidence, it is imperative that we observe more BH XRB sources in the hard state.\\

\ixpe observations of Type-1 Active Galactic Nuclei (AGNs) may also help constrain the properties of the X-ray corona, since current understanding suggests that the hard X-rays produced by AGNs originate from a corona similar to that found in X-ray binaries. The PA measurements of NGC~4151 \citep{Gianolli_2023,Gianolli_2024}, IC~4329A \citep{ingram2023xraypolarisationseyfert1} and \mbox{MCG-05-23-16} \citep{Marinucci_2022} have been found to align with the radio jet or ionisation cone, suggesting again that the corona is extended in the disc plane. The observed PD values are also rather high ($\sim 3$~per cent for IC~4329A and $\sim 5$~per cent for NGC~4151), given that Type-1 AGNs are thought to be viewed from a fairly low inclination according to the unification scheme.

% In IC4329 A, an outflowing corona was suggested to account for the low observed PD.\\
%studies have shown that AGN have the same properties of BH XRB with scaled up properties. 
%These are currently the only two BH XRB sources that have been observed in the hard state, so in order to determine whether the similarities between these sources is significant or coincidence, we need to observe more sources.\\

\igr is a transient BH XRB first discovered in 2003 by \integral \citep{Kuulkers2003}, and is of particular interest due to its extraordinary variability behaviour. While \igr exhibits all the spectral states of a typical XRB \citep{Capitanio_2012}, it also displays 10 different exotic variability states \citep{Altamirano2011,Court2017,Capitanio_2012,Wang2024}. This makes it one of two systems known to exhibit this behaviour, the other source being \grs, which has exhibited 14 different variability states \citep{Belloni2000}, 7 of which in common with \igr, including a heartbeat-like pattern \citep{Wang2024}. These variability states were first hypothesised from observations of \grs as radiation pressure instabilities deriving from its high-Eddington accretion rates, but because \igr has demonstrated the same variability states at a flux estimated to be $\sim$20--30 times lower, this model is now being questioned. For example, the presence of substantial wind outflows from the accretion disk might significantly influence these instability patterns, potentially stabilizing or suppressing the observed heartbeat oscillations \citep{Janiuk_2015}.

% The source parameters of IGR J17091$-$3624, including inclination, are not well constrained. Different studies using reflection spectrum modelling have yielded values from $i\sim24 - 45$, with higher values suggested to account for the source's apparent low flux \citep{Xu_2017,Wang2018igr,Wang2024}.\\
% \\
Here we report the results of the \ixpe observation of \igr in the hard state, taken in March 2025. We also discuss two shorter simultaneous observations by the \textit{Nuclear Spectroscopic Telescope ARray} (\nustar). 
In Section~\ref{sec:data}, we detail the data reduction procedure. In Section~\ref{sec:results}, we present the polarimetric analysis as well as spectro-polarimetric fits. In Section~\ref{sec:discuss} we discuss our results and in Section~\ref{sec:summary} we summarise our conclusions.\\
%It is therefore essential that more XRBs are observed with polarimetry missions in order to determine the prevailing model.
% the trajectory of a photon after a scattering is most likely to be in the plane perpendicular to its electric polarisation vector.
%This is because scattered photons are polarised perpendicular, and an asymmetric corona will have more photons in the largest plane axis due to there being a higher scattering optical depth, so the net polarisation is perpendicular to the longest axis of the asymmetric corona. 
%where a higher polarisation is indicative of a higher inclination angle, due to the more asymmetric view. 
%Cygnus X-1, a persistent XRB was observed to have a polarisation degree of ~4\%, which was considerably higher tham that expected from
%1. intro: hard state bh xrb, corona wtf, xraypol ixpe break degeneracies, cygx1 4 percent was high :was it misaligned (implies lots of low pol systems) or outlfow (high pol is common)?
%swift 4 percent also: only two, both 4 percent, coincidence?
%igr j17091 canonical states plus crazy grs1915 states
%maybe agn?
%here we report on a hard state observation by ixpe  of igr
%which had been calibrated using the standard \ixpe pipeline

\section{Observation and data reduction}
\label{sec:data} 
\subsection{\ixpe data}

\igr was observed by \ixpe on  2025 March 7--10  for a total elapsed time of $\sim 300$~ks (obsID: 0450201). We measured a mean count rate of $\sim 1$ count~s$^{-1}$ with a useful exposure time of $\sim160$~ks.

\ixpe is a joint NASA/Italian Space Agency mission launched on 2021 December 9 from the Kennedy Space Center. It consists of three identical gas pixel detector units (DUs) that record the spatial, energy, timing, and polarimetric information from each event within a 2--8~keV energy band. Specifications and observatory details can be found in \citet{Weisskopf2022}.

\begin{figure}
\centering
%    \begin{minipage}{0.49\textwidth}
%        \centering
        %\vspace{10mm}
\includegraphics[width=0.95\linewidth,trim=2.0cm 4.0cm 1.0cm 3.0cm,clip=true]{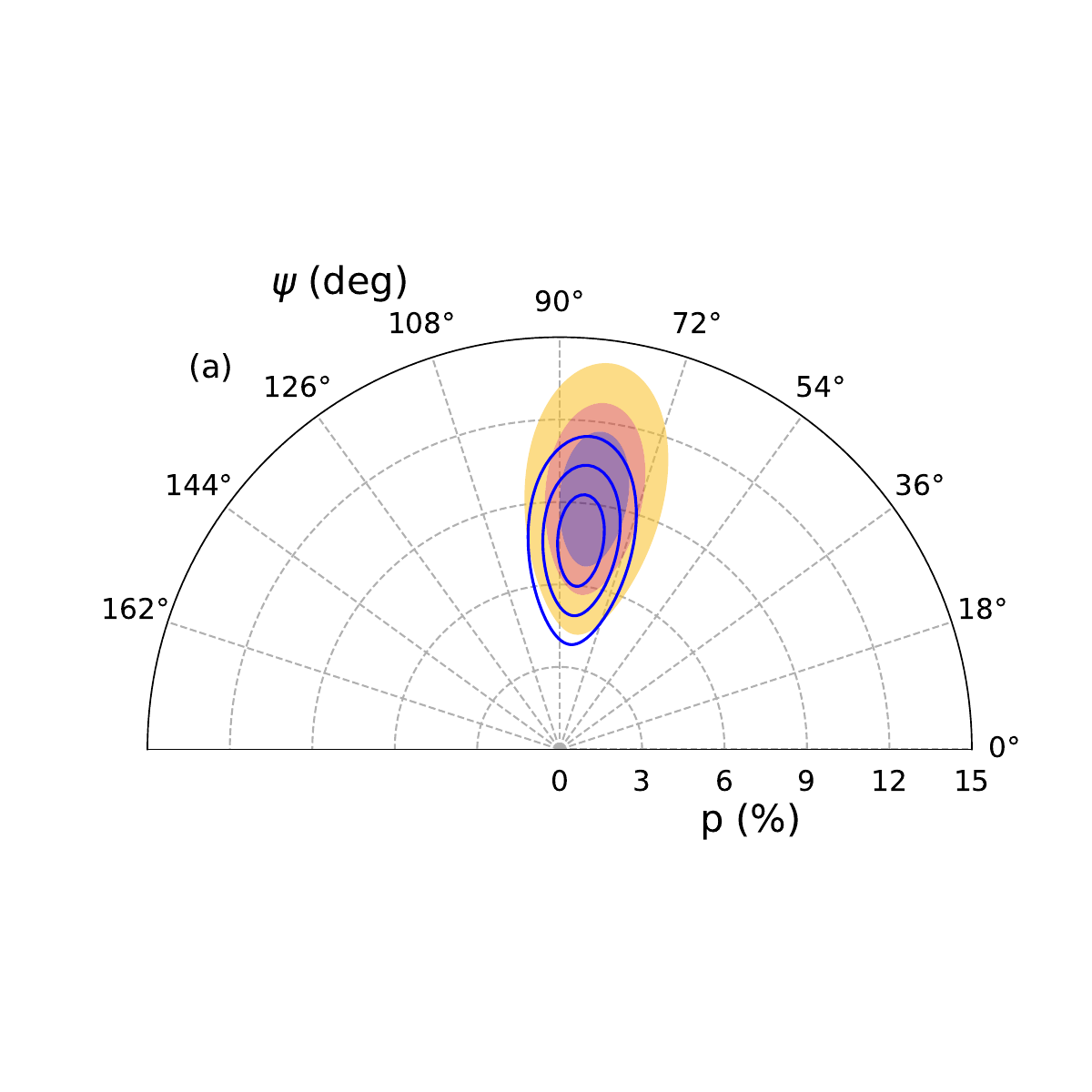}
 %   \end{minipage}%
 %   \hfill
%    \begin{minipage}{0.49\textwidth}
%        \centering
%        \vspace{5mm}      
\includegraphics[width=\linewidth]{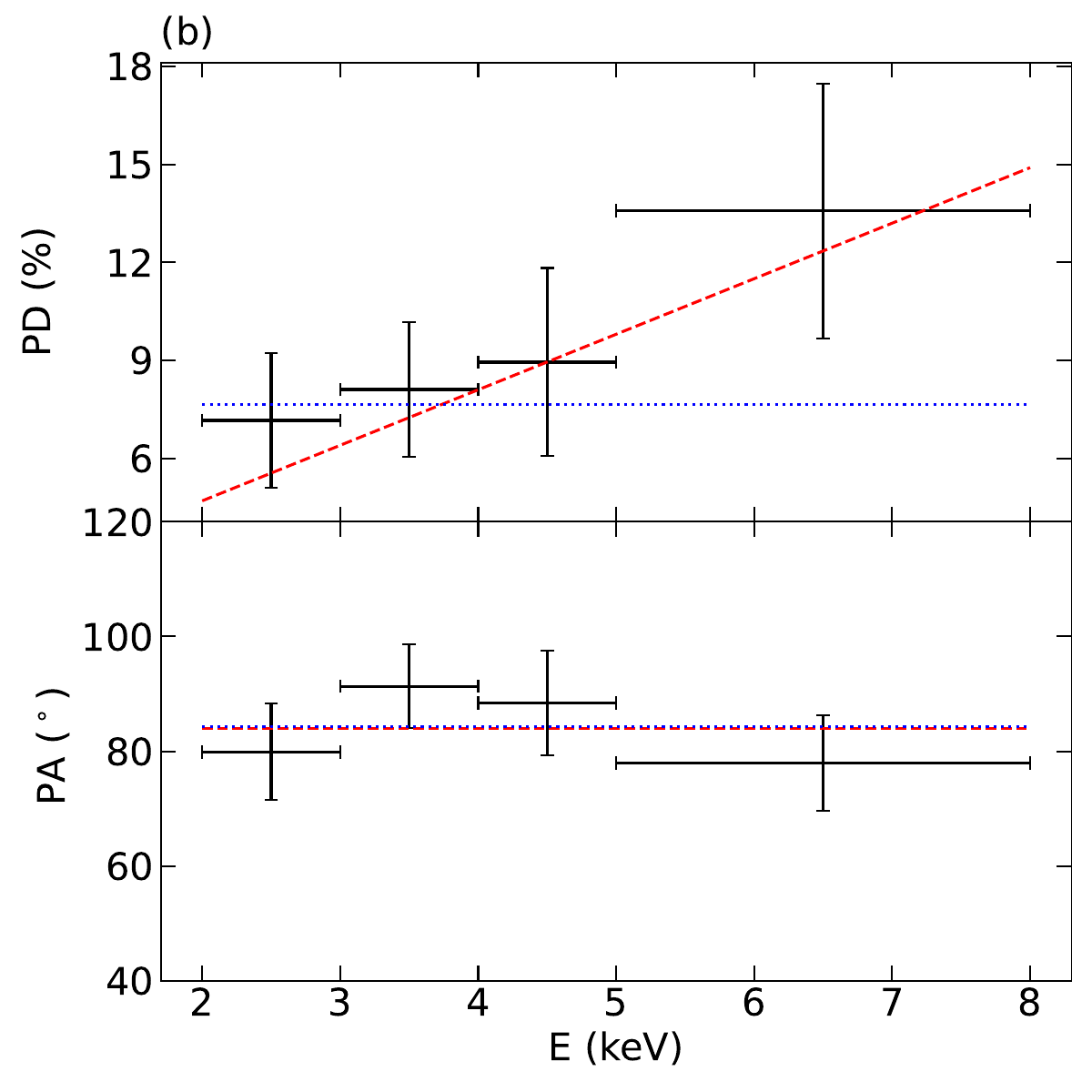}
     %\label{fig:image2}
%    \end{minipage}
\caption{Polarimetric properties of \igr.
(a) Constraints on the PD and PA (measured East of North) in the 2--8 keV band. The contours represent the confidence levels at 68, 90 and 99 per cent. The filled colour contours are the results of the unweighted, model-independent \texttt{pcube} algorithm. The blue contours are the results of an unweighted spectro-polarimetric analysis within \texttt{xspec}. (b) Energy dependence of the observed PD (top) and PA (bottom) with $1\sigma$ error bars as measured by the \texttt{pcube} algorithm. The red dashed lines represent the best-fitting energy dependent spectro-polarimetric model (row 4 of Table \ref{tab:1}), where the PD increases with energy while the PA remains constant. The blue dashed lines represent the best fitting constant polarisation model.}
\label{fig:PDPAe}
\end{figure}
%Left, Bottom, Right, Top

We downloaded the cleaned Level 2 event files for each DU \citep{Costa2001} from the High Energy Astrophysics Science Archive Research Center (HEASARC\footnote{\url{https://heasarc.gsfc.nasa.gov/docs/ixpe/archive/}}). We used \texttt{ds9} to define a circular source region centred on the source with a radius of 60\arcsec, and a background region of an annulus centred on the source with an inner radius of 150\arcsec\ and an outer radius of 300\arcsec. Using the latest version of \textsc{ixpeobssim} (v31.0.3; \citealt{BALDINI2022101194}) we used the \texttt{xpselect} tool to create source and background fits files filtered for these regions. For our model-independent analysis, we used the \texttt{pcube} algorithm to calculate the Stokes parameters for the source and background regions, and subtract the background Stokes parameters from the source. For the spectro-polarimetric analysis, we used the \texttt{PHA} algorithm to extract Stokes $I$, $Q$ and $U$ as a function of energy channel for the source and background region of each detector unit, employing the most recent calibration database files associated with the latest version of \textsc{ixpeobssim}, using an energy bin width of 0.016~keV. Throughout our analysis, we do not employ track weightings, and the PA is defined as East of North.

%This method is model-independent and does not apply weightings.

%Pcube analysis? -- Model independent but no weighting
%pha and xspec -- model dependent, weightings applied
\subsection{\nustar data}

\nustar \citep{Harrison2013} observed the source twice during the \ixpe exposure (obsID 81002342008 and 81002342010). The observations started on 2025 March 7 at 14:00 UTC and on March 8 at 23:30 UTC, respectively, and each had a 20~ks exposure time. 
We used the \texttt{nupipeline} and \texttt{nuproducts}
routines distributed from the \texttt{NuSTARDAS v2.1.4} release with the calibration files \texttt{20250310} to produce the event files and the energy spectra for both Focal Plane Modules (FPMs).
We chose an extraction region of 100\arcsec\ centred on the peak of the source photon counts. The same size region positioned as close as possible to the source was chosen for the background extraction. 
After preliminary analysis suggested that the spectra were broadly consistent between the two observations, we combined the FPMA and FPMB spectra from the two observations, as well as their backgrounds and response files, using \texttt{addascaspec} from the \textsc{ftools} package. This was done to increase the signal for plotting purposes only. For a detailed analysis, it is best to treat the different observations and FPMs separately.

\section{Results}
\label{sec:results}
\subsection{Model-independent analysis}

We first measured the 2--8 keV polarisation using the \textsc{ixpeobssim} algorithm \texttt{pcube}.  We found ${\rm PD}=9.1\pm 1.6$ per cent and ${\rm PA}=83\degr\pm 5\degr$ ($1\sigma$ uncertainties), and we plot the statistical confidence contours in Fig.~\ref{fig:PDPAe}a (colour filled contours).
%PD(%)= [9.138706]
%dPD(%)= [1.6410847]
%MDP(%)= [4.9782104]
%PA(deg)= [82.751305]
%dPA(deg)= [5.144449]

To check if there is a dependence of polarisation properties with energy, we used the same algorithm to calculate the PD and PA over a span of energy bins. The results can be seen in Fig. \ref{fig:PDPAe}b, with the PA showing no apparent energy dependence. The PD shows a possible increase with energy, which we test in the following subsection.

\begin{figure}
    \centering
    \includegraphics[width=\linewidth,trim=0.0cm 0.0cm 1.5cm 1.0cm,clip=true]{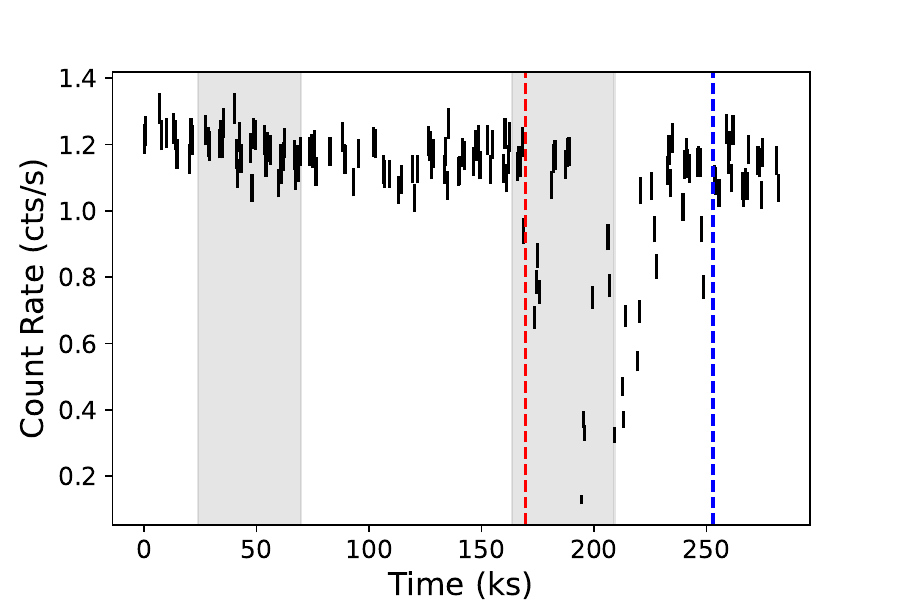}
    \caption{Light curve in counts per second of \igr as observed by \ixpe (black bars). The counts from each DU have been summed. The dip begins at the horizontal dashed red line, and ends at the horizontal blue dashed line. The shaded grey regions represent the two simultaneous \nustar  observations. The error bars are derived from Poisson statistics.}
    \label{fig:lc}
\end{figure}

% \subsection{Dips}
During the \ixpe observation, \igr experienced a dip in the X-ray flux reaching as low as $\sim$0.2~count~s$^{-1}$ over all DUs, starting $\sim$150 ks into the observation, and lasting for $\sim$50~ks. We plot the light curve of the total observation in Fig.~\ref{fig:lc}, where the dip can clearly be observed. To test any effects that this may have on the observed polarisation, we calculated the polarisation properties excluding the events within the dip, again utilising the \texttt{pcube} algorithm. We found PD=$8.4\pm1.8$~per cent and PA=$86\degr\pm 6\degr$, which is consistent with the results calculated for the entire observation. We therefore conclude that due to the low number of counts contributed from the dip to the total observation, it does not have a significant effect on the observed polarisation.
%and we include the dip for all subsequent analysis. 

%Left, Bottom, Right, Top

\subsection{Spectro-polarimetric analysis}
\label{sec:specpol} 

Using \textsc{xspec} v12.14.1 \citep{Arnaud1996}, we simultaneously fitted Stokes $I$, $Q$ and $U$ with several basic models representing different polarisation properties. The polarisation results and fit statistics for each model can be found in Table~\ref{tab:1}. We include a multiplicative \texttt{constant} component in each fit to account for cross-calibration between the DUs.  
We first fit the data with a model consisting of  \texttt{diskbb} \citep{Mitsuda1984} and \texttt{powerlaw} components, representing a multi-temperature black-body accretion disc and a power law with specific photon flux $\propto E^{-\Gamma}$. 
The interstellar absorption was accounted for using the model  \texttt{tbabs} with the relative abundances from \citet{Wilms2000}.

We first tested models with a constant, energy independent polarisation using model \texttt{polconst} (rows 2--3 in Table \ref{tab:1}). For the model including both disc and powerlaw components, i.e. \texttt{tbabs*polconst*(diskbb+powerlaw)}, we measure PD=$7.6 \pm 1.8$~per cent and PA$=84\degr\pm 7\degr$ with a fit statistic $\chi^2$/d.o.f=1320/1334. This inferred polarisation is consistent with that measured using \texttt{pcube}, and we plot the resulting confidence contours in Fig.~\ref{fig:PDPAe}a (blue hollow contours).

\begin{figure}
    \centering
    \includegraphics[width=1\linewidth]{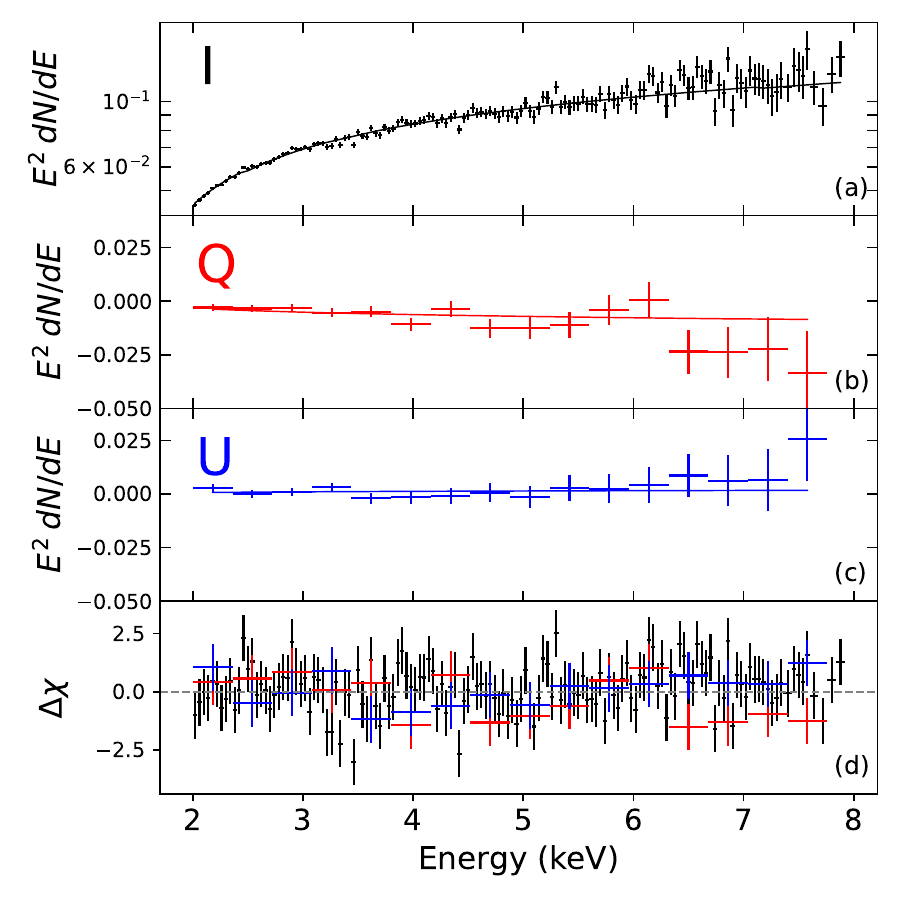}
    \caption{Spectral energy distribution of the constant polarisation model (row~3 of Table~\ref{tab:1}). 
    For plotting purposes only, we group the three DUs and employ energy rebinning to achieve a target signal to noise of 6 by binning no more than 9 channels (i.e. the \texttt{xspec} command \texttt{setplot rebin 6 9}).
    (a) Best-fitting spectral model for Stokes $I$ (black line) with unfolded data (black crosses). (b) Best-fitting spectral model (red line) with unfolded Stokes $Q$ (red crosses). (c) Best-fitting spectral model (blue line) with unfolded Stokes $U$ (blue crosses). (d) Contributions to fit statistic $\chi$. $dN/dE$ is in units of photons ${\rm cm}^{-2}~{\rm s}^{-1}~{\rm keV}^{-1}$. }
    \label{fig:eeuf}
\end{figure}

We then fit the model without the \texttt{diskbb} component, i.e. \texttt{tbabs*polconst*(powerlaw)}, which returned a fit statistic of $\chi^2$/d.o.f=1326/1334. We find that the polarisation properties between the two constant polarisation models are consistent, and that the disc is only detected in the flux spectrum with a statistical confidence of $1.8~\sigma$ according to an F-test.
%the model including the \texttt{diskbb} component is only preferred over the model without it by $1.8~\sigma$ according to an F-test and so is not statistically significant.
We plot the unfolded Stokes $I$, $Q$ and $U$ and the contributions to the fit statistic of the constant polarisation model excluding the disc component in Fig.~\ref{fig:eeuf}. 

We then fit models to check for an energy dependence of the polarisation properties by replacing \texttt{polconst} with \texttt{pollin}, which assumes a linear dependence of PD and PA with photon energy (rows 3--5 of Table~\ref{tab:1}). The linear relationship is described as $P(E)=P_1+(E/{\rm keV} -1)\times P_{\rm slope}$ where $P(E)$ is the energy dependent polarisation property (i.e. either PD or PA), $P_1$ is the polarisation property measured at 1 keV and $P_{\rm slope}$ is the gradient. We perform the fit under two different assumptions. The first assumption is that only the PD is variable with energy where the PA remains constant. This is achieved by freezing the slope parameter of PA to 0. In the second case, we assume that both PD and PA are energy dependent. We test these assumptions with (\texttt{tbabs*pollin*(diskbb+powerlaw)}) and without (\texttt{tbabs*pollin*(powerlaw)}) a disc component.
We again find that the polarisation results are consistent between models with and without the \texttt{diskbb} component, and so from here we continue our analysis considering the simpler models containing only the \texttt{powerlaw} component.

To test if a polarisation linearly dependent on energy is preferred to a constant polarisation, we perform F-tests between the \texttt{polconst} model and the \texttt{pollin} model under the two aformentioned assumptions. The first test assuming only PD varies yields $F=2.93$, meaning the variable PD is only preferred by $1.7\sigma$. Similarly for an F-test where PD and PA are free to vary, this model is only preferred to the constant polarisation model by $1.3\sigma$. Although there is a hint of positive trending of PD with energy, no significant energy dependence of polarisation properties has been detected. 
%For plotting purposes only, we group the three DUs and employ energy rebinning to achieve a target signal to noise of 4 by binning no more than 4 channels (i.e. the \texttt{xspec} command \texttt{setplot rebin 4 4}).

%To test the effect of the dip on the extracted Stokes $I$, we attempted to fit a basic spectral model, \texttt{tbabs*const*powerlaw}, to events associated with the dip, but found that the quality of the data were too poor to achieve an acceptable fit. We then fit the events associated with the dip removed and found a power-law index $\Gamma=1.68 \pm0.03$ and $N_{\rm H}=(1.5\pm0.1)\times 10^{22}$~cm$^{-2}$ for a fit statistic $\chi^2$/d.o.f=411/447 which is consistent with the values found from using the entire observation. We therefore conclude that the dip does not have a significant effect on our spectro-polarimetric analysis.

\begin{table*}
   \centering
\begin{tabular}{lllllc ccc c}
        \toprule\toprule
        {Model} & \multicolumn{2}{c}{PD (\%)} & \multicolumn{2}{c}{PA (deg)} & $N_{\rm H}$ ($10^{22}$ cm$^{-2}$) & $\Gamma$ & $kT_{\rm bb}$ (keV) & $\chi^2$/dof \\
        \midrule
        \texttt{pcube} & \multicolumn{2}{c}{$9.1 \pm 1.6$} & \multicolumn{2}{c}{$83 \pm 5$} & - & - & - & - \\
        \midrule
        \texttt{polconst*tbabs*(diskbb+po)} & \multicolumn{2}{c}{$7.6\pm 1.8$} & \multicolumn{2}{c}{$84\pm 7$} & $2.3^{+0.9}_{-0.8}$ & $1.56^{+0.15}_{-0.28}$ & $0.48^{+0.51}_{-0.11}$ & 1320/1332 \\
        \midrule
        \texttt{polconst*tbabs*po} & \multicolumn{2}{c}{$7.6 \pm 1.8$} & \multicolumn{2}{c}{$84 \pm 7$} & $1.39^{+0.15}_{-0.15}$ & $1.58^{+0.04}_{-0.04}$ & - & 1326/1334 \\
        \midrule
        \multirow{2}{*}{\texttt{pollin*tbabs*(diskbb+po)}} & $\rm PD_1$ & $3^{+5}_{-3}$ & $\rm PA_1$ & $84 \pm 7$ & \multirow{2}{*}{$2.0^{+1.8}_{-0.7}$} & \multirow{2}{*}{$1.52^{+0.18}_{-0.17}$} & \multirow{2}{*}{$0.54^{+0.47}_{-0.16}$} & \multirow{2}{*}{1318/1331} \\
        & $\rm PD_{\text{slope}}$ & $1.6^{+1.5}_{-1.6}$ & $\rm PA_{\text{slope}}$ & $0.0$ & & & & \\
        \midrule
        \multirow{2}{*}{\texttt{pollin*tbabs*po}} & $\rm PD_1$ & $3^{+5}_{-3}$ & $\rm PA_1$ & $84 \pm 7$ & \multirow{2}{*}{$1.39^{+0.15}_{-0.15}$} & \multirow{2}{*}{$1.59^{+0.02}_{-0.04}$} & \multirow{2}{*}{-} & \multirow{2}{*}{1323/1333} \\
        & $\rm PD_{\text{slope}}$ & $1.7^{+1.5}_{-1.6}$ & $\rm PA_{\text{slope}}$ & $0.0$ & & & & \\
        \midrule
        \multirow{2}{*}{\texttt{pollin*tbabs*(diskbb+po)}} & $\rm PD_1$ & $3^{+5}_{-3}$ & $\rm PA_1$ & $90^{+10}_{-16}$ & \multirow{2}{*}{$2.0^{+1.8}_{-0.7}$} & \multirow{2}{*}{$1.52^{+0.16}_{-0.17}$} & \multirow{2}{*}{$0.54^{+0.47}_{-0.16}$} & \multirow{2}{*}{1317/1330} \\
        & $\rm PD_{\text{slope}}$ & $1.8^{+1.4}_{-1.7}$ & $\rm PA_{\text{slope}}$ & $-2^{+4}_{-3}$ & & & & \\
        \midrule
        \multirow{2}{*}{\texttt{pollin*tbabs*po}} & $\rm PD_1$ & $3^{+5}_{-3}$ & $\rm PA_1$ & $90^{+17}_{-18}$ & \multirow{2}{*}{$1.39^{+0.15}_{-0.15}$} & \multirow{2}{*}{$1.59^{+0.04}_{-0.04}$} & \multirow{2}{*}{-} & \multirow{2}{*}{1323/1332} \\
        & $\rm PD_{\text{slope}}$ & $1.8^{+1.4}_{-1.7}$ & $\rm PA_{\text{slope}}$ & $-1.6^{+5.2}_{-1.6}$ & & & & \\
        \bottomrule
    \end{tabular}
    \caption{Model fitting results showing polarisation degree (PD), polarisation angle (PA), hydrogen column density ($N_{\rm H}$), power-law index ($\Gamma$), peak disc blackbody temperature ($kT_{\rm bb}$), and chi-square per degree of freedom ($\chi^2$/dof).}
    \label{tab:1}
\end{table*}

We note that, although the \texttt{pcube} and \texttt{polconst} measurements of PD are consistent within $1\sigma$ confidence, the \texttt{polconst} estimate is systematically lower (see e.g. Fig.~\ref{fig:PDPAe}a). This is likely because the PD is increasing with energy, as has been observed for other higher signal to noise datasets \citep[e.g.][]{ingram2024trackingxraypolarizationblack}, and is hinted at here but with low statistical confidence. The \texttt{polconst} model has, by design, constant PD and so the best-fitting PD value is weighted towards the lower energies, for which there are more counts. The \texttt{pcube} algorithm, in contrast, calculates the polarisation properties averaged over the 2--8~keV bandpass (weighted by flux). To illustrate this, we plot in Fig.~\ref{fig:PDPAe}b the best-fitting PD and PA as a function of energy for our  \texttt{polconst} (blue) and \texttt{pollin} (red) models. We see that, as expected, \texttt{pollin} agrees with the \texttt{pcube} points, whereas the \texttt{polconst} PD is slightly lower (whilst still within $2~\sigma$ uncertainties). We therefore adopt the \texttt{pcube} values as the best estimate of the 2--8~keV PD and PA.

\subsection{State classification}

\begin{figure}
    \centering\includegraphics[width=\linewidth,trim=0.3cm 0.35cm 0.2cm 0.3cm,clip=true]{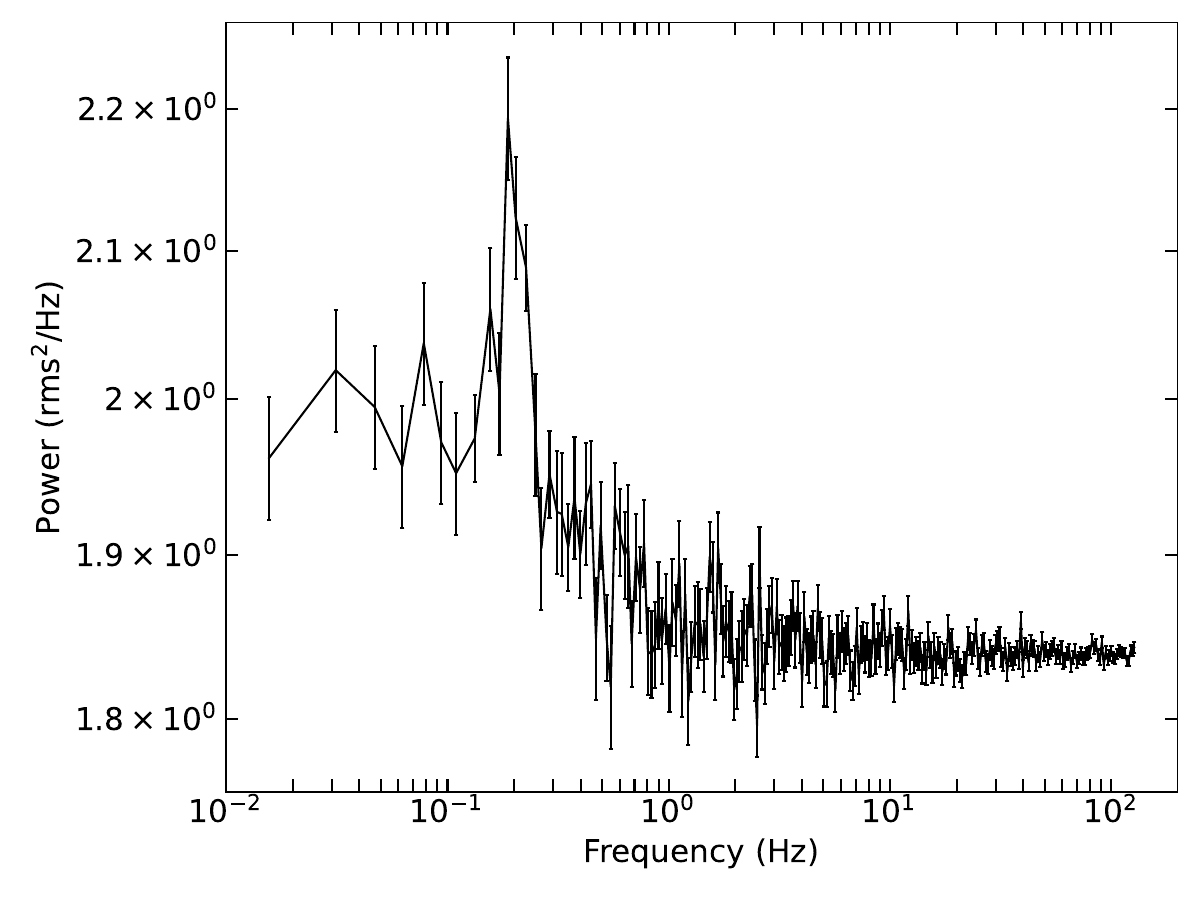}
    \caption{Power spectrum of the \ixpe observation in the 2--8~keV band, calculated in fractional rms normalisation, where Poisson noise has not been subtracted. The input light curve was binned in 1/256~s time bins and the power spectrum was averaged over a segment size of 64~s. Logarithmic rebinning was applied (with the geometric rebinning parameter set to $f=0.2$).}
    \label{fig:ps}
\end{figure}

For all models, we find that the hydrogen column density is in the range $N_{\rm H} \approx (1.4-2.3) \times 10^{22}~{\rm cm}^{-2}$ \citep[consistent with previous measurements;][]{Wang2024} and the power-law index is $\Gamma \approx 1.6$ (see Table~\ref{tab:1}). These values are consistent with that seen in \cite{Wang2024}. This photon index indicates that \igr was in the hard state during the \ixpe observation.
% Through our spectro-polarimetric fits, we also constrain the hydrogen column density $N_{\rm H}$ and the power law index $\Gamma$ in order to inspect the spectral state of the source. The results can be found in Table \ref{tab:1}. We find that $\Gamma \approx 1.6$, which is consistent with that of XRBs observed in the hard state and nH$\sim1.4-2.3$, which is not large enough to greatly suppress the soft emission.
To further consolidate the hard state classification, we also studied the timing properties. We created a light curve from all DUs with time bins of 1/256 s duration, and filtered the events such that they coincided with when all the DUs were in a good time interval (GTI). We then created a power spectrum averaged over a segment length of $T_{\rm seg}=64$~s using the \textsc{Stingray} python package \citep{Huppenkothen_2019}. It is clear from Fig.~\ref{fig:ps} that a Type-C quasi-periodic oscillation (QPO) is present at $\sim 0.2$ Hz. Based on this combination of spectral and timing evidence, we conclude that the source is in a hard state and that the X-ray flux is dominated by the corona.

\begin{figure}
    \centering
    \includegraphics[width=0.85\linewidth]{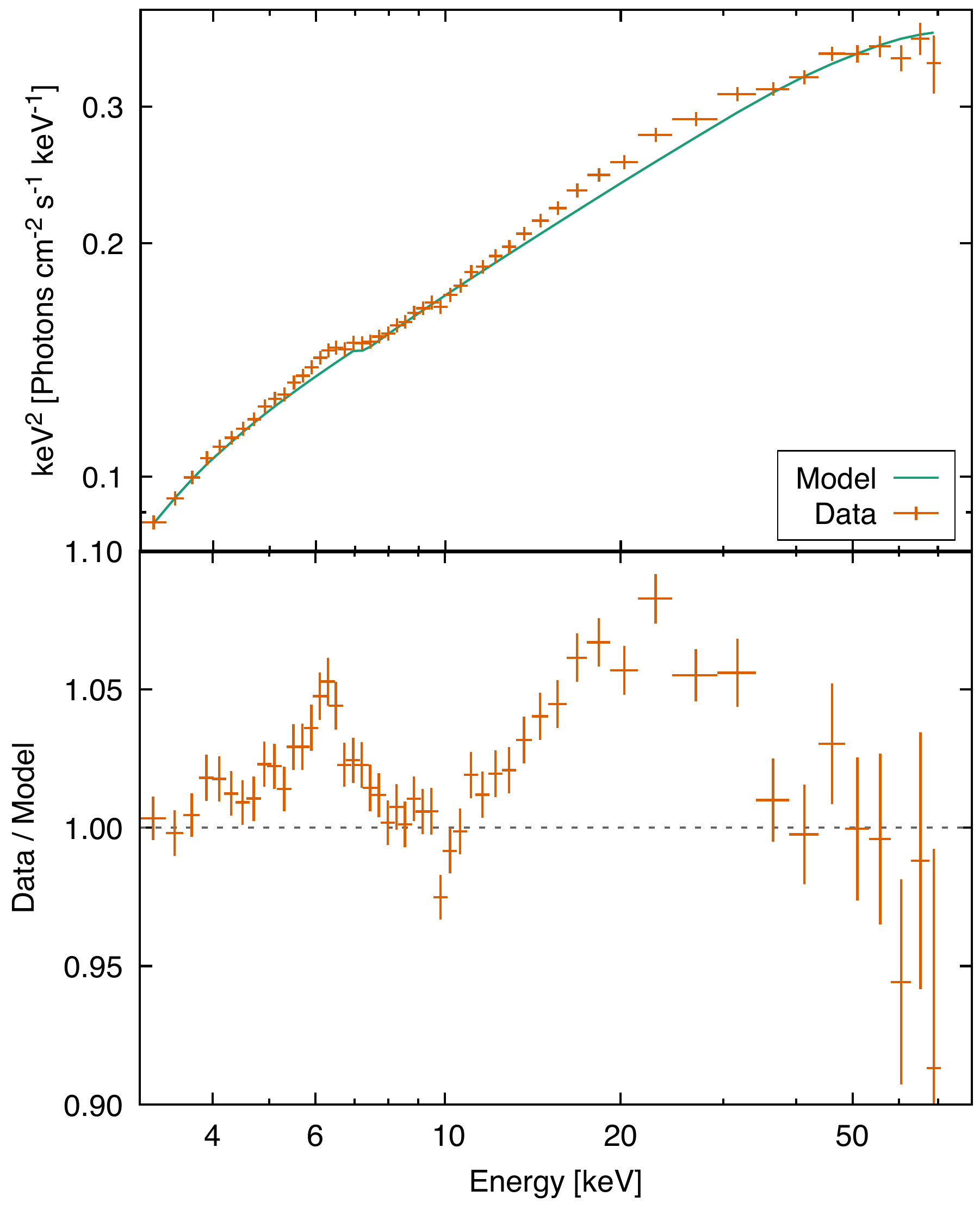}
    \caption{Unfolded \nustar energy spectrum fitted with the model \texttt{tbabs*nthComp} (top panel) and the corresponding residuals (bottom panel). FPMA and FPMB data are grouped for plotting purposes.}
    \label{fig:nustar_spectrum}
\end{figure}

We also considered the \nustar observations taken during the \ixpe exposure (see the grey shaded regions in Fig. \ref{fig:lc}).
% performed at the beginning of the \ixpe exposure. 
We find that the energy spectrum in the 3--79~keV \nustar energy range is dominated by the Comptonised emission, such that a disc component is not required to fit the spectrum. 
% For this reason, a thermal disk component is not required to fit the \nustar data. 
Figure~\ref{fig:nustar_spectrum} shows the results of fitting the combined spectrum of the two \nustar observations with the model \texttt{tbabs * nthComp}, where \texttt{nthcomp} is a thermal Comptonisation model \citep{Zdziarski1996}. We fix the hydrogen column density to $N_{\rm H} = 1.1 \times 10^{22}~{\rm cm}^{-2}$, the seed blackbody photon temperature to $kT_{\rm bb}=0.1$~keV, and the coronal electron temperature to $kT_{\rm e} = 40$~keV. This electron temperature is motivated by preliminary fits, but we note that a reliable constraint on $kT_{\rm e}$ requires a very detailed analysis beyond the scope of this paper. The measured photon index of $\Gamma=1.607 \pm 0.002$ is consistent with what we measure with \ixpe (see Table~\ref{tab:1}). The residuals to this fit clearly reveal an iron line and Compton hump indicative of reflection. The presence of a clear Compton hump contributes to our hard state classification, since the Compton hump is much weaker than the iron line for a soft illuminating spectrum \citep{Garcia2022}. The relative weakness of the reflection features (the residuals in the \ixpe band are below 5 per cent) confirms that the 2--8~keV flux is dominated by Comptonised emission with a small (likely $\sim$10--20~per cent based on preliminary fits) contribution from reflection.
We leave a detailed spectro-polarimetric analysis jointly considering the \ixpe and \nustar observations for a follow up paper. 

% Left Bottom Right Top

% \begin{figure}
%     \centering
%     \includegraphics[width=1\linewidth]{Plots/igr_j17091_NuSTAR_spectrum_nth_fit.pdf}
%     \caption{Unfolded \nustar energy spectra (FPMA in black and FPMB in red) fitted with the model \texttt{TBabs*nthComp} (top panel) and the corresponding residuals (bottom panel). The seed photons of the Comptonisation component are produced by a disk-blackbody model with disk temperature fixed to 0.1~keV.}
%     \label{fig:nustar_spectrum}
% \end{figure}

\section{Discussion}
\label{sec:discuss}

We have presented an \ixpe~observation of \igr~in the hard state and find that the 2--8~keV polarisation is ${\rm PD}= 9.1 \pm 1.6$~per cent and ${\rm PA} =83\degr \pm 5\degr$. We are not able to compare the PA with the orientation of the jet in \igr because it has never been spatially resolved by radio observations. Additionally, simultaneous MeerKAT observations do not significantly detect radio polarisation (Russell et al., in prep).
As such, we are unable to make any inference on the relative orientation of the corona, but for the purposes of this discussion, we will assume that it is extended in the disc plane as is indicated by the polarisation aligning with the jet for all X-ray polarisation detections of hard state BH XRBs and Seyfert-1 galaxies with resolved jets \citep{Krawczynski_2022,Gianolli_2023,Veledina2023, ingram2023xraypolarisationseyfert1}.
Simultaneous {Very Large Telescope} (VLT) optical polarimetric observations show an increasing PD with frequency ($2.9 \pm 0.5$ and $1.0 \pm 0.3$ per cent in the $R$ and $I$ bands, respectively; Baglio et al., in prep.), with a PA of $100\degr \pm 5\degr$ and $102\degr \pm 8\degr$, consistent with the \ixpe PA within $2\sigma$. This behaviour aligns with recent findings on BH XRBs in their hard state, where the optical PA has been observed to be similar to the PA in the X-rays, for example, in Cyg~X-1 \citep{Krawczynski_2022,Kravtsov2023}, Swift J1727.8$-$0127 \citep{Veledina2023}  and GX~339$-$4 \citep{Mastroserio2025}, and the position angle of the radio jet in V404~Cyg \citep{Kosenkov2017} and MAXI~J1820+070 \citep{Veledina2019}. Similarly, also for \igr we favour a scenario in which the optical radiation originates in the outer regions of the disc and is polarised through Thomson scattering in the disc atmosphere or wind.

The observed 2--8~keV PD of $\approx$9~per cent is significantly higher than what has been measured thus far for other sources in corona-dominated states (e.g. $\sim$4 per cent for both Cyg~X-1 and Swift~J1727$-$1613; \citealt{Krawczynski_2022,Veledina2023}). 
Because the PD strongly depends on inclination, it is likely that \igr is more highly inclined than the other sources. However, the inclination of \igr is not well constrained. Reflection spectroscopy modelling yields estimates of fairly low inclination \citep[$i \sim 30\degr-40\degr$;][]{Xu_2017}, whereas the presence of dipping in the light curve \citep[e.g. our Fig.~\ref{fig:lc} and][]{Pahari_2013} and wind signatures in the spectrum \citep{Wang2024,Parra2024_winds,Ponti2012}, coupled with the absence of eclipses instead indicates that the inclination is in the range $60\degr \lesssim i \lesssim 80\degr$ . The high amplitude of the Type-C QPO also points towards high inclination \citep{Motta2015}, consistent with the high inclination indicated by the high X-ray PD we observe here.

\begin{figure}
\centering
\includegraphics[width=\linewidth,trim=0.5cm 0.0cm 1.0cm 1.0cm,clip=true]{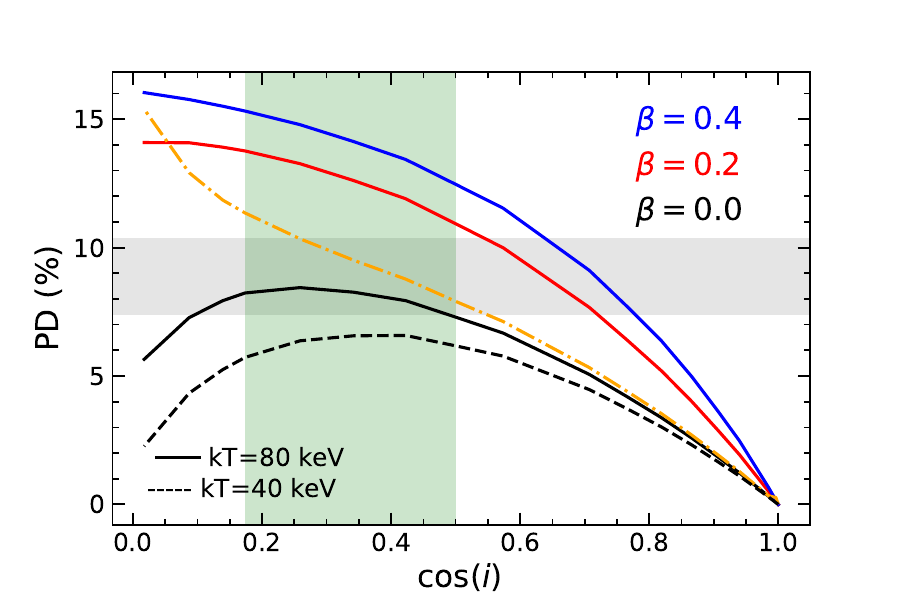}
\caption{Predicted PD for a Comptonisation model in a slab-corona as a function of the cosine of the inclination for different bulk velocities. The black, blue and red lines represent bulk velocities of $\beta=v/c=0$, $0.2$ and $0.4$, respectively. 
The solid and dashed lines correspond to coronal temperatures of $kT_{\rm e}=80$ and $40$~keV, respectively. The orange dot-dashed line represents the PD influenced by scattering in a wind.
The coronal optical depth was adjusted to produce spectral slope $\Gamma=1.6$.
The grey horizontal shaded area represents the limits on the observed PD measurement. The vertical green shaded region is the most likely range of inclination angles.}
\label{fig:pdcosi}
\end{figure}

A PD of 9 per cent is high even for a high inclination source. 
% high, particularly in comparison to other sources, including AGN. One potential model to explain this high result is if the corona has a relativistic bulk outflow velocity, which would in turn increase the observed PD due to relativistic aberration. This was first proposed in \cite{Poutanen_2023} and has been suggested in other observations such as \cite{Krawczynski_2022} and  \cite{ingram2023xraypolarisationseyfert1}.
To demonstrate this, we show in Fig. \ref{fig:pdcosi} the predicted PD as a function of inclination angle for a slab corona with seed photons provided from the side by a truncated disk (see Fig. 1c of \citealt{Poutanen_2023}). This geometry is chosen to maximise the predicted PD (see Fig. 2 of \citealt{Poutanen_2023}). We therefore note that these predictions should be seen as an upper limit of the expected PD, since any change to the assumed geometry (e.g. large coronal opening angles, significant inhomogeneities of the photosphere shape, etc.) would reduce the predicted PD. We use the analytic code \texttt{compps} (\citealt{Poutanen1996}; see also \citealt{Veledina_2022}) and assume a 
% truncated disc and a slab coronal geometry with
power-law index of $\Gamma=1.6$
%, an electron temperature of $kT_{\rm e}=80$~keV
and a disc seed photon temperature of $kT_{\rm bb}=0.1$ keV (typical for the hard-state sources and optimal to maximise the PD).
% These values are typical of an XRB in the hard state, and the electron temperature and seed photon temperature were chosen to maximise the observed polarisation. 
The grey shaded area represents the $1~\sigma$ confidence interval of the observed PD. The black lines represent the case where the electrons in the corona have no bulk velocity ($\beta=v/c=0$), only random thermal velocities. For the dashed line, we assume an electron temperature of $kT_{\rm e}=40$ keV, which is consistent with the fairly low value of $kT_{\rm e}$ yielded by our initial fits to the \nustar data. We see that the observed PD is above this line for all viewing angles. However, without a very detailed spectral analysis, $kT_{\rm e}$ is highly uncertain. We therefore also plot calculations assuming $kT_{\rm e}=80$ keV as solid lines, which maximises the predicted PD. We see that under this assumption, the $\beta=0$ model is consistent with the measured PD within uncertainties for inclinations in the range $60\degr \lesssim i \lesssim 80\degr$ as highlighted by the green shaded region.

A non-zero bulk outflow velocity, $\beta>0$, of coronal electrons would increase the predicted PD due to relativistic aberration \citep{Poutanen_2023}. We therefore plot examples of $\beta=0.2$ and $0.4$ in Fig.~\ref{fig:pdcosi}. We see that such values of the outflow velocity can reproduce the observed PD for much lower inclination angles. If the inclination of \igr is $i \lesssim 60\degr$ and the high PD results from bulk motion, this implies that other, higher inclination sources will be observed in future to have even higher PD.
% , implying that other BH XRBs may in the future be observed to have even higher PD. 

Scattering in a disc wind adds another mechanism capable of affecting the polarisation properties \citep{Tomaru2024,Nitindala_2025}, and \igr is well known for strong wind signatures, albeit only in its exotic variability classes \citep{Wang2024}. The resulting PD depends on the optical depth of wind, its characteristic opening angle, angular distribution and polarisation of the incident X-ray emission.
In Fig.~\ref{fig:pdcosi} we show as the orange dot-dashed line an example for the PD boost by scattering in the wind for the case of incident radiation corresponding to Comptonization in the slab corona of $kT_{\rm e}=80$~keV producing a power-law spectrum with   $\Gamma=1.6$.
We assumed the wind density profile following a Gaussian with opening angle $\alpha_{\rm w}=20\degr$ \citep{Nitindala_2025} and a mid-plane Thomson optical depth of $\tau_{\rm T}=1$.

The same process could affect the optical polarisation as well.
%Although the faintness of the optical counterpart prevents detections of "cold" wind signatures with optical spectroscopy, the source is expected to show wind signatures in optical and infrared in the current spectral state, and this cold wind component is likely to influence the polarization as well. 
Our VLT polarimetric observations show an optical PD that increases slightly with frequency and a PA parallel to the X-ray PA measured with \ixpe, suggesting that the polarisation originates from scattering within the plane of the accretion disc. This scattering could occur either in the atmosphere of the viscously heated optically thick accretion disc itself or in a disc wind, and distinguishing between these two scenarios is not possible. A more detailed analysis will be presented in a forthcoming paper focused on the optical polarisation properties of the source (Baglio et al. in prep.).

%Cristina follow-up?

%Another mechanism that would boost PD is scattering in an optically thick wind \citep{Nitindala_2025}.

% We see that the PD typically increases with inclination angle as the corona becomes more symmetric to our field of view. With the grey shaded region as the error measurement on the observed PD, we see that even in conditions set to maximise PD, it is high for $\beta=0$. This could suggest that a bulk velocity $\beta>0$ is required, where further observations of hard state XRBs would facilitate the study of this model.

Reflection can also influence the observed PD, since the reflected emission can be more polarised than the directly observed coronal emission. The reflected emission is expected to be polarised perpendicular to the disc plane, thus aligning with the polarisation of the direct coronal emission (which we are assuming to align with the jet). In this case, the overall 2--8~keV PD is given by
% as we may be measuring polarisation due to the reprocessing of photons from the corona in the disc, and so it is important to be taken into account. Simultaneous \nustar observations took place alongside \ixpe observations, and from preliminary spectral fits, we can show that the contribution of reflection is not particularly strong. Although not very well constrained, the fit indicates that the electron temperature is relatively low, which would favour a lower observed PD. The overall PD of an XRB can be described by
\begin{equation}
    {\rm PD}=[(1-R)\times {\rm PD}_{\rm cor}]+R\times {\rm PD}_{\rm ref},
\end{equation} 
where $R$ is the fraction of the 2--8~keV flux that is contributed by reflection, ${\rm PD}_{\rm cor}$ is the PD of the corona and ${\rm PD}_{\rm ref}$ is the PD of the reflection component. The maximum expected PD of the reflection component is $\sim 20$~per cent \citep{matt1993b,Poutanen1996b,Podgorny2023}, and a typical hard state 2--8~keV reflected fraction is $R\sim$0.1--0.2 \citep[e.g.][]{Krawczynski_2022}, which appears to be consistent with our \nustar data (Fig.~\ref{fig:nustar_spectrum}). Therefore our observation of PD$\approx9$~per cent gives a lower limit for the PD of the directly observed emission of ${\rm PD}_{\rm cor}\gtrsim 6$~per cent, which is still reasonably high compared with the expectations of thermal Compton scattering with no bulk velocity. More rigorous exploration of the physics of the corona will be provided by detailed spectro-polarimetric fits that jointly consider the \ixpe and \nustar data, which we leave to future work.

\section{Conclusions}
\label{sec:summary}

We have obtained the first X-ray polarisation measurement for the BH XRB \igr. The time- and energy-averaged ${\rm PD} =9.1\pm 1.6$~per cent and ${\rm PA} =83\degr \pm 5\degr$ ($1\sigma$ errors) was measured with statistical confidence $5.2\sigma$ using the model-independent \texttt{pcube} algorithm. We find no statistically significant dependence of the PD or PA on energy.
% The dip in flux does not influence the observed polarisation properties.
From the shape of the energy spectrum,
% From constraints on nH and $\Gamma$ through our fits,
and the presence of a $\sim$0.2~Hz Type-C QPO in the power spectrum, we confirm that the source was observed in the hard state where the X-ray flux is dominated by the corona. Without the orientation of the radio jet we cannot confirm the orientation of the corona with respect to the disc, but we do find that the X-ray PA is consistent with the $I$ and $R$ band polarization within $2~\sigma$ confidence, as has previously been found for other BH XRBs \citep{Krawczynski_2022,Mastroserio2025}.
% but expect it to be parallel from previous observations of other XRBs and AGN.
The very high observed PD requires \igr to have a favourable coronal geometry and to be viewed from a favourable angle for it to be explained by standard thermal Comptonisation models. If we are instead viewing from a lower inclination angle, the PD could have been boosted by the electrons in the corona having a mildly relativistic bulk outflow velocity, and/or by scattering in an optically thick disc wind. If this is true, we expect in future to observe other hard state BH XRBs with higher inclination angles that exhibit even higher PD.

\section*{Acknowledgements}

% The Acknowledgements section is not numbered. Here you can thank helpful
% colleagues, acknowledge funding agencies, telescopes and facilities used etc.

M.E. and A.I. acknowledge support from the Royal Society. P.O.P. acknowledges financial support from the French Space National Agency (CNES) and the National Center of Scientifc Research (CNRS) via its "Action Thématique" PEM. T.D.R. and M.C.B. are INAF IAF research fellows. G.Mattand S.B. acknowledge financial support by the Italian Space Agency (Agenzia Spaziale Italiana, ASI) through the contract ASI-INAF-2022-19-HH.0. B.D.M. acknowledges support via a Ram\'on y Cajal Fellowship (RYC2018-025950-I), the Spanish MINECO grants PID2023-148661NB-I00, PID2022-136828NB-C44, and the AGAUR/Generalitat de Catalunya grant SGR-386/2021. M.D. and M.G. thank GACR project 21-06825X for the support and institutional support from RVO:67985815. The work of G.Matt and L.M. is partially supported by the PRIN 2022 - 2022LWPEXW - “An X-ray view of compact objects in polarized light”, CUP C53D23001180006.
A.V. acknowledges support from the Academy of Finland grant 355672. Nordita is supported in part by NordForsk.
%%%%%%%%%%%%%%%%%%%%%%%%%%%%%%%%%%%%%%%%%%%%%%%%%%
\section*{Data Availability}
\ixpe data are publicly available from the HEASARC data archive (\url{https://heasarc.gsfc.nasa.gov)}.



\bibliographystyle{mnras}
\bibliography{main} % if your bibtex file is called example.bib

%\begin{thebibliography}{99}
%\end{thebibliography}

%%%%%%%%%%%%%%%%%%%%%%%%%%%%%%%%%%%%%%%%%%%%%%%%%%

%%%%%%%%%%%%%%%%% APPENDICES %%%%%%%%%%%%%%%%%%%%%

\appendix

%%%%%%%%%%%%%%%%%%%%%%%%%%%%%%%%%%%%%%%%%%%%%%%%%%

% Don't change these lines
\bsp	% typesetting comment
\label{lastpage}
\end{document}